\documentclass[conference]{IEEEtran}
\IEEEoverridecommandlockouts

\usepackage{cite}
\usepackage{amsmath,amssymb,amsfonts}
\usepackage{textcomp}

\usepackage{algorithm}
\usepackage{algpseudocode}

\usepackage{listings}
\usepackage{graphicx}
\usepackage{pythonhighlight}
\usepackage{times}
\usepackage[utf8]{inputenc}
\usepackage[colorlinks,linkcolor=blue]{hyperref}
\usepackage{float}
\usepackage{endnotes}
\usepackage{colortbl}
\usepackage[numbers]{natbib} 
\usepackage{caption}
\usepackage{xcolor}

\makeatletter
\renewcommand\subsubsection{
  \@startsection{subsubsection}{3}{0em}
    {1ex plus 0.5ex minus 0.2ex}
    {0.7ex plus .5ex minus 0ex}
    {\normalfont\normalsize\itshape}
}
\makeatother

\newcommand{\subsubsectionstar}[1]{
  \vspace{0.7ex plus 0.5ex minus 0ex}
  \noindent\hspace{0em}\textit{#1}\par
}

\makeatletter
\providecommand{\columnbreak}{\par\vfill\break}
\makeatother

\begin{document}

\title{High-Performance Parallelization of Dijkstra's Algorithm Using MPI and CUDA}

\author{Boyang Song\\
\textit{Department of Computer Science}\\
\textit{Cornell University}\\
Ithaca, NY, USA\\
bs834@cornell.edu}

\maketitle

\begin{abstract}
    This paper investigates the parallelization of Dijkstra's algorithm for computing the shortest paths in large-scale graphs using MPI and CUDA. The primary hypothesis is that by leveraging parallel computing, the computation time can be significantly reduced compared to a serial implementation. To validate this, I implemented three versions of the algorithm: a serial version, an MPI-based parallel version, and a CUDA-based parallel version. Experimental results demonstrate that the MPI implementation achieves over 5x speedup, while the CUDA implementation attains more than 10x improvement relative to the serial benchmark. However, the study also reveals inherent challenges in parallelizing Dijkstra's algorithm, including its sequential logic and significant synchronization overhead. Furthermore, the use of an adjacency matrix as the data structure is examined, highlighting its impact on memory consumption and performance in both dense and sparse graphs.
\end{abstract}

\section{Introduction}
The hypothesis of this project is that implementing the Dijkstra's algorithm using the MPI / CUDA will significantly reduce the computation time for finding the shortest path in large-scale graphs compared to its serial version.

The problem explored by the project is the impact of using MPI / CUDA parallel Dijkstra's algorithm on performance for large-scale graphs, especially computation time and scalability.

I implement the serial Dijkstra's algorithm, MPI Dijkstra's algorithm, and CUDA Dijkstra's algorithm respectively, and explore the project by comparing the performance of these three implementations.

Ultimately, my MPI parallel implementation achieved more than \textbf{5x} performance improvements running Dijkstra's algorithm on large-scale graphs, and the CUDA implementation achieved more than \textbf{10x} performance improvements compared to the serial version.

\section{Key prior work}
Original Dijkstra's Algorithm\cite{DIJKSTRA1959}: The foundational work by E.W. Dijkstra, which introduces the algorithm for finding the shortest paths between nodes in a graph, serves as the basis for my implementation.

\section{Empirical methodology}
Performance is measured as the execution time of graphs of the same scale under three implementations.

The input is Edge Lists. The program writes them to the adjacency matrix. Since this process is the same in the three implementations and does not involve parallelization of Dijkstra's algorithm, no costs are included. The performance cost range includes the full range of costs required for parallelization \footnote{Use \textit{{\#include \textless chrono\textgreater}
} for timing. Both of parallel dijkstra's algorithm (MPI and CUDA) include Synchronization cost}

\begin{itemize}
    \item Serial: dijkstra's algorithm cost
    \item MPI: MPI communication cost, parallel dijkstra's algorithm cost, MPI Gather cost
    \item CUDA: GPU memory cost, GPU communication cost,  parallel dijkstra's kernel algorithm cost
\end{itemize}

\subsection{Serial implementation}
A simple serial implementation of Dijkstra's algorithm. This is the starting point for parallel optimization of Dijkstra's algorithm. Subsequent MPI parallelism and CUDA parallelism are based on this serial version.

The program will convert the input Edge Lists into an adjacency matrix according to the parameters\footnote{Add the parameter \textit{-w} during runtime to fill as a Directed graph. Unless otherwise stated, undirected graphs will be used in this article.}.

\begin{minipage}{.5\textwidth}
\centering
\[
\begin{array}{c|cccc}
  & 0 & 1 & 2 & 3 \\ \hline
0 & 0 & 2 & 4 & \infty \\
1 & 2 & 0 & 1 & 3 \\
2 & 4 & 1 & 0 & 5 \\
3 & \infty & 3 & 5 & 0 \\
\end{array}
\]
\text{Undirected graph adjacency matrix}
\end{minipage}

\begin{minipage}{.5\textwidth}
\centering
\[
\begin{array}{c|cccc}
  & 0 & 1 & 2 & 3 \\ \hline
0 & 0 & 2 & 4 & \infty \\
1 & \infty & 0 & 1 & 3 \\
2 & \infty & \infty & 0 & 5 \\
3 & \infty & \infty & \infty & 0 \\
\end{array}
\]
\text{Directed graph adjacency matrix}
\end{minipage}

\begin{figure}[H] 
\centering
\includegraphics[width=0.4\textwidth]{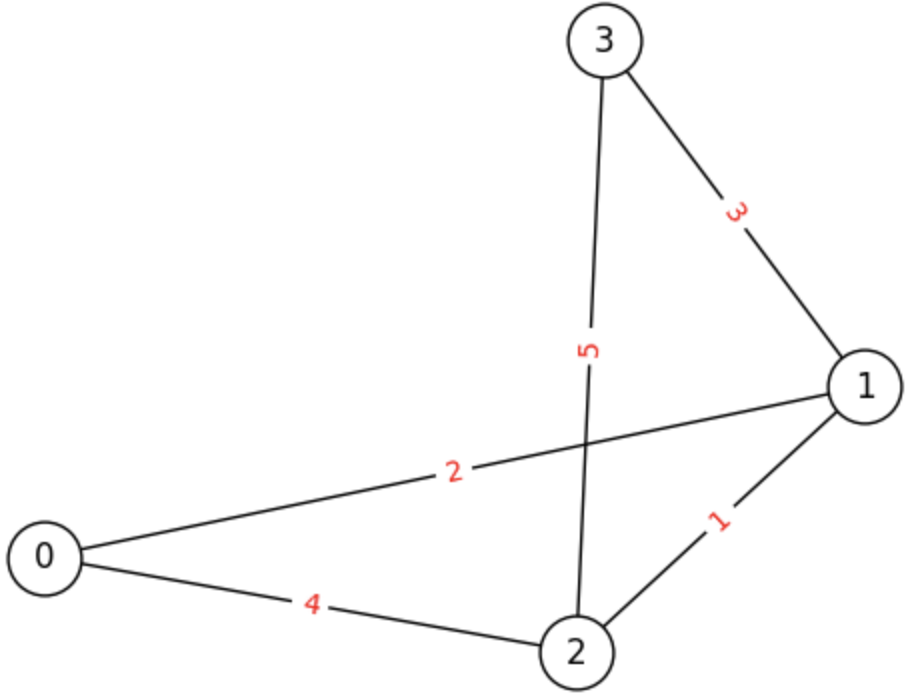}
\caption{Undirected graph}
\end{figure}

\begin{algorithm}[H]
\captionsetup{labelformat=empty}
\caption{\textbf{Algorithm 1 } Serial Dijkstra's Algorithm}
\begin{algorithmic}[1]
\Function{Dijkstra}{$G, s$}
    \State $n \gets \text{number of nodes in } G$
    \State \textbf{let} $dist[]$ be an array of size $n$
    \State \textbf{let} $pred[]$ be an array of size $n$
    \State \textbf{let} $visited[]$ be an array of size $n$
    \State Initialize $dist[i] \gets \infty$ and $visited[i] \gets \text{false}$ for all $i$
    \State $dist[s] \gets 0$
    \For{$i \gets 0$ to $n-1$}
        \State $u \gets \text{node with min } dist[u] \text{ that is not visited}$
        \State $visited[u] \gets \text{true}$
        \For{each neighbor $v$ of $u$}
            \If{\raggedright $\text{not } visited[v]\text{ and } G[u, v] \neq \infty \text{ and } dist[u] + G[u, v] < dist[v]$}
                \State $dist[v] \gets dist[u] + G[u, v]$
                \State $pred[v] \gets u$
            \EndIf
        \EndFor
    \EndFor
\EndFunction
\end{algorithmic}
\end{algorithm}

\subsection{MPI Parallel implementation}

Based on the serial Dijkstra's Algorithm implementation, MPI is used for process communication to achieve high-performance parallel Dijkstra's Algorithm.

\subsubsection{MPI preparation}

\begin{algorithm}[H]
\captionsetup{labelformat=empty}
\caption{MPI preparation}
\begin{algorithmic}[1]
\State MPI\_Init(\&argc, \&argv)
\State comm = MPI\_COMM\_WORLD
\State MPI\_Comm\_rank(comm, \&my\_rank)
\State MPI\_Comm\_size(comm, \&procs)
\end{algorithmic}
\end{algorithm}

\subsubsection{Data reading and distribution - distributes the data required by each procs}

The main proc (rank == 0) completes the initial reading, and then distributes the data to different procs.

Distribute data according to column division. Considering the characteristics of Dijkstra's Algorithm, 1D division method is more appropriate.

In order to handle the situation where the number of columns is not divisible by the number of procs.

\begin{algorithm}[H]
\captionsetup{labelformat=empty}
\caption{Calculate Padded Vertices Number}
\begin{algorithmic}[1]
    \If{$procs > n$}
        \State $padded\_n \gets procs$ \Comment{Set padded\_n directly to procs}
    \Else
        \State $remainder \gets n \% procs$ \Comment{Calculate the remainder of n divided by procs}
        \If{$remainder \neq 0$}
            \State $adjustment \gets procs - remainder$ \Comment{Calculate necessary adjustment}
        \Else
            \State $adjustment \gets 0$ \Comment{No adjustment needed if n is already a multiple of procs}
        \EndIf
        \State $padded\_n \gets n + adjustment$ \Comment{Adjust n to the nearest multiple of procs}
    \EndIf
\end{algorithmic}
\end{algorithm}

In large adjacency matrices with many nodes ($>$100), the performance penalty due to padding is minimal. This trick makes it easier to balance loads.

I use 3 procs to process a 4 node matrix (4 columns) for example.

\begin{minipage}{.5\textwidth}
\centering
\[
\begin{array}{c|cccccc}
  & 0 & 1 & 2 & 3 & \cellcolor{gray!20}N & \cellcolor{gray!20}N \\ \hline
0 & 0 & 2 & 4 & \infty & \cellcolor{gray!20}\infty& \cellcolor{gray!20}\infty\\
1 & 2 & 0 & 1 & 3 & \cellcolor{gray!20}\infty& \cellcolor{gray!20}\infty\\
2 & 4 & 1 & 0 & 5 & \cellcolor{gray!20}\infty& \cellcolor{gray!20}\infty\\
3 & \infty & 3 & 5 & 0 & \cellcolor{gray!20}\infty& \cellcolor{gray!20}\infty\\
\cellcolor{gray!20}N & \cellcolor{gray!20}\infty & \cellcolor{gray!20}\infty & \cellcolor{gray!20}\infty & \cellcolor{gray!20}\infty & \cellcolor{gray!20}\infty & \cellcolor{gray!20}\infty\\
\cellcolor{gray!20}N & \cellcolor{gray!20}\infty & \cellcolor{gray!20}\infty & \cellcolor{gray!20}\infty & \cellcolor{gray!20}\infty & \cellcolor{gray!20}\infty & \cellcolor{gray!20}\infty\\
\end{array}
\]
\text{Padded matrix}
\end{minipage}

\begin{minipage}{.5\textwidth}
\centering
\[
\begin{array}{  c|
  >{\columncolor{cyan!20}}c
  >{\columncolor{cyan!20}}c
  |>{\columncolor{yellow!20}}c
  >{\columncolor{yellow!20}}c|
  >{\columncolor{magenta!20}}c
  >{\columncolor{magenta!20}}c}
  & 0 & 1 & 2 & 3 & N & N \\ \hline
0 & 0 & 2 & 4 & \infty & \infty& \infty\\
1 & 2 & 0 & 1 & 3 & \infty& \infty\\
2 & 4 & 1 & 0 & 5 & \infty& \infty\\
3 & \infty & 3 & 5 & 0 & \infty& \infty\\
N & \infty & \infty & \infty & \infty & \infty & \infty\\
N & \infty & \infty & \infty & \infty & \infty & \infty\\
\end{array}
\]
\text{Abstract division}
\end{minipage}

After that, use the following method to distribute the data:

The main proc (rank == 0) distributes data, for itself, it simply copies the data to its own local buffer, for other processes, use \textit{MPI\_Send} to send the corresponding portion of the data to each process.

Other process receive data, use \textit{MPI\_Recv} receive their respective data.

\begin{algorithm}[H]
\captionsetup{labelformat=empty}
\caption{Simplified MPI Data Distribution}
\begin{algorithmic}[1]
    \State MPI\_Bcast(\&n, 1, MPI\_INT, 0, comm) \Comment{Broadcast total number of nodes}
    \State loc\_n = n / procs \Comment{Compute local number of nodes}
    \If{my\_rank == 0}
        \For{proc = 0 to procs - 1}
            \If{proc == 0}
                \State \textit{Copy data to local buffer}\Comment{Simply copy the data to its own local buffer}
            \Else
                \State MPI\_Send(data to proc, loc\_n, MPI\_INT, proc, 0, comm) \Comment{Send submatrix (column)}
            \EndIf
        \EndFor
    \Else
        \State MPI\_Recv(data from 0, loc\_n, MPI\_INT, 0, 0, comm, MPI\_STATUS\_IGNORE) \Comment{Receive submatrix (column)}
    \EndIf
\end{algorithmic}
\end{algorithm}

\subsubsection{Parallel Dijkstra's algorithm}
After the distribution of the graph data among the processes, each process performs the following steps to execute the parallel Dijkstra's algorithm.

\begin{algorithm}[H]
\captionsetup{labelformat=empty}
\caption{\textbf{Algorithm 2 } Simplified Parallel Dijkstra's Algorithm using MPI}
\begin{algorithmic}[1]
\Function{DijkstraMPI}{$loc\_Adj, loc\_dist, n, comm$}
    \State $my\_rank \gets \text{MPI\_Comm\_rank(comm)}$
    \State Initialize $loc\_dist[]$ for local nodes
    \State $min\_dist \gets \infty$
    \State $loc\_u \gets -1$

    \For{$i \gets 0 \text{ to } n-1$}
        \If{$loc\_dist[i] < min\_dist$ and $loc\_dist[i]$ is local minimum}
            \State $min\_dist \gets loc\_dist[i]$
            \State $loc\_u \gets i$
        \EndIf
    \EndFor

    \State Perform $\text{MPI\_Allreduce}$ to find global minimum distance \Comment{key step}
    \State Update local distances based on global minimum
\EndFunction
\end{algorithmic}
\end{algorithm}

The traditional serial Dijkstra's algorithm selects the nearest node that is not currently visited at each step and then updates the distance of its adjacent nodes. In the MPI parallel version, the adjacency matrix of the graph is divided into multiple blocks, and each process processes a piece of data, that is, each process is responsible for calculating the shortest path from some nodes to all nodes.

Through \textit{MPI\_Allreduce}, all processes jointly decide which node is the current global shortest distance node. This operation ensures that each process has a global view, that is, all processes know which is the current shortest distance node.

\subsubsection{MPI Gather }
Summarize results scattered across different processes.

\subsection{CUDA Parallel implementation}
Based on the serial Dijkstra's Algorithm implementation, CUDA distributes the adjacency matrix in the GPU memory and then performs parallel  Dijkstra's Algorithm.

\subsubsection{Memory allocation and data transfer }
Allocate memory by \textit{cudaMalloc} on the GPU for data that supports operations, and copy the data by \textit{cudaMemcpy} from the host to the GPU memory.

\subsubsection{ Kernel function Call and Calling logic}

\begin{algorithm}[H]
\captionsetup{labelformat=empty}
\caption{\textbf{Algorithm 3 } Simplified Parallel Dijkstra's Algorithm using CUDA}
\begin{algorithmic}[1]
\Function{DijkstraCUDA}{$n, Adj, dist, pred$}
    \State\raggedright Allocate GPU memory for $d\_Adj, d\_dist, d\_pred, d\_updated$
    \State Copy $Adj, dist, pred$ from host to GPU
    \State Setup $blocks$ and $threads$ for kernel execution
    \Repeat
        \State $anyUpdated \gets false$
        \State Reset $d\_updated$ array on device to false
        \State Execute $dijkstra\_kernel$ on GPU \Comment{dijkstra\_kernel will be explained in detail below}
        \State Synchronize device
        \State $anyUpdated \gets$ Thrust::reduce(Thrust::device, d\_updated, d\_updated + n, false, Thrust::logical\_or) \Comment{Use Thrust to perform a reduction operation on the GPU and check if there is a true value}
    \Until{$not \ anyUpdated$}
    \State Copy $dist$ and $pred$ from device to host
    \State Free GPU resources
\EndFunction
\end{algorithmic}
\end{algorithm}

Set the number of blocks and threads required for CUDA kernel function execution, execute \textit{dijkstra\_kernel} in a loop until there are no more updates.

Checking whether there is an update is necessary. Here, the \textit{reduce} function of the \textbf{Thrust} library is used to quickly check whether there are any true values in the \textit{d\_updated} in GPU, which avoids the cost of exchanging data between the host and the GPU. The \textbf{Thrust} library makes it easier and more efficient to implement complex data processing in CUDA.

The \textit{logical\_or} operation checks whether at least one element is true. If there is any true in the range, the final result will be true; otherwise, the result will be false.

\subsubsection{Kernel function}

This kernel function is executed in parallel, and each thread is responsible for updating the shortest path of a node, allowing the GPU to process multiple nodes at the same time, greatly improving computing efficiency.

The logic of algorithm implementation is similar to that of serial, but there are many differences in implementation details to adapt to GPU parallel requirements. The main difference is that \textit{atomicMin} is used to ensure the correctness and thread safety when multiple threads update shared data (dist).

\begin{algorithm}[H]
\captionsetup{labelformat=empty}
\caption{\textbf{Algorithm 4 } Dijkstra's Parallel Algorithm Kernel using CUDA}
\begin{algorithmic}[1]
\Function{DijkstraKernel}{$n, Adj, dist, pred, updated$}
    \State $tid \gets blockIdx.x \times blockDim.x + threadIdx.x$
    \If{$tid \geq n$}
        \State \Return
    \EndIf
    \For{$v = 0$ to $n-1$}
        \State $weight \gets Adj[tid \times n + v]$
        \If{$weight \neq INF$ \textbf{and} $dist[tid] \neq INF$}
            \State $new\_dist \gets dist[tid] + weight$
            \If{$new\_dist < dist[v]$}
                \State $old\_dist \gets atomicMin(\&dist[v], new\_dist)$\Comment{\textit{atomicMin} is used to ensure the correctness and thread safety}
                \If{$new\_dist < old\_dist$}
                    \State $pred[v] \gets tid$
                    \State $updated[v] \gets true$
                \Comment{If dist[v] is updated by the new path length, update the predecessor node of node v to tid, and mark node v as updated in the updated array.}
                \EndIf
            \EndIf
        \EndIf
    \EndFor
\EndFunction
\end{algorithmic}
\end{algorithm}

\section{Result}
The experiments in this project were conducted on the Perlmutter\footnote{Compute Nodes\\ 
CPU: 1 $\times$ AMD EPYC 7763 (64 cores, 2.45 GHz)\\
GPU: 4 $\times$ NVIDIA A100 (80GB).}.

To evaluate the performance gains of MPI and CUDA parallelizations, three implementations (Serial, MPI, and CUDA) were used to process graphs of various scales and densities.

\subsection{Graphs used for performance evaluation}
Three implementations are used to process the following graphs of different scales/densities respectively. MPI is tested under different number of processes to analyze its performance.

\begin{table}[H]
\centering
\begin{tabular}{|l|c|c|}
\hline
Graph ID & \# of Nodes & \# of Edges \\
\hline
Graph 10-45 & 10 & 45 \\
Graph 100-4950 & 100 & 4950 \\
Graph 1000-499500 & 1000 & 499500 \\
Graph 2000-1899500 & 2000 & 1899500 \\
\hline
\end{tabular}
\caption{Dense Graphs}
\end{table}
\begin{table}[H]
\centering
\begin{tabular}{|l|c|c|}
\hline
Graph ID & \# of Nodes & \# of Edges \\
\hline
Graph 10-30 & 10 & 30 \\
Graph 100-300 & 100 & 300 \\
Graph 1000-3000 & 1000 & 3000 \\
Graph 2000-6000 & 2000 & 6000 \\
Graph 10000-30000 & 10000 & 30000 \\
Graph 20000-60000 & 20000 & 60000 \\
Graph 40000-120000 & 40000 & 120000 \\
\hline
\end{tabular}
\caption{Spare Graphs}
\end{table}

\begin{enumerate}
    \item Compare the performance of three implementations.
    \item Compare the differences between dense and sparse graphs.
    \item For MPI parallelism, compare the differences between different procs (Scaling).
\end{enumerate}
\subsection{Performance of three implementations}
For the convenience of display, I used 32 procs MPI to compare with serial and CUDA parallelism. More results of MPI parallelism (dashed lines and semi-transparent) will be displayed in \ref{sec:MPI parallelism Details}. The performance of dense graphs and sparse graphs are the same, which I will show in \ref{sec:Dense and Sparse Graphs}. In the plots, I use the data of sparse graphs for comparison.

\begin{figure}[h]
    \centering
    \includegraphics[width=0.9\columnwidth]{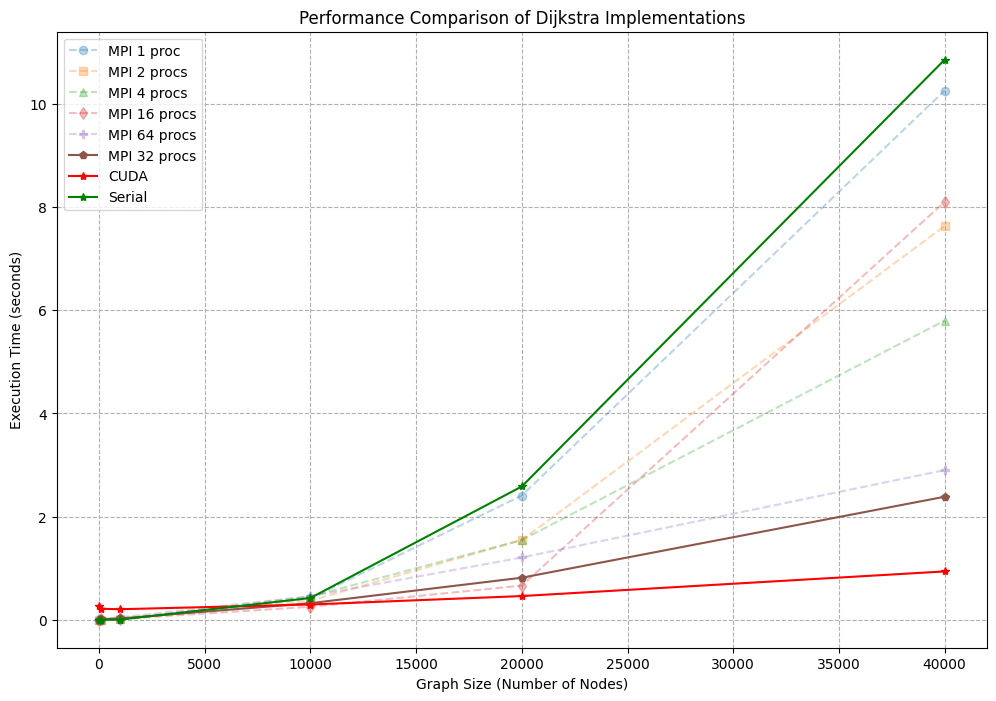}
    \caption{Performance (linear scale)}
\end{figure}

\begin{figure}[h]
    \centering
    \includegraphics[width=0.9\columnwidth]{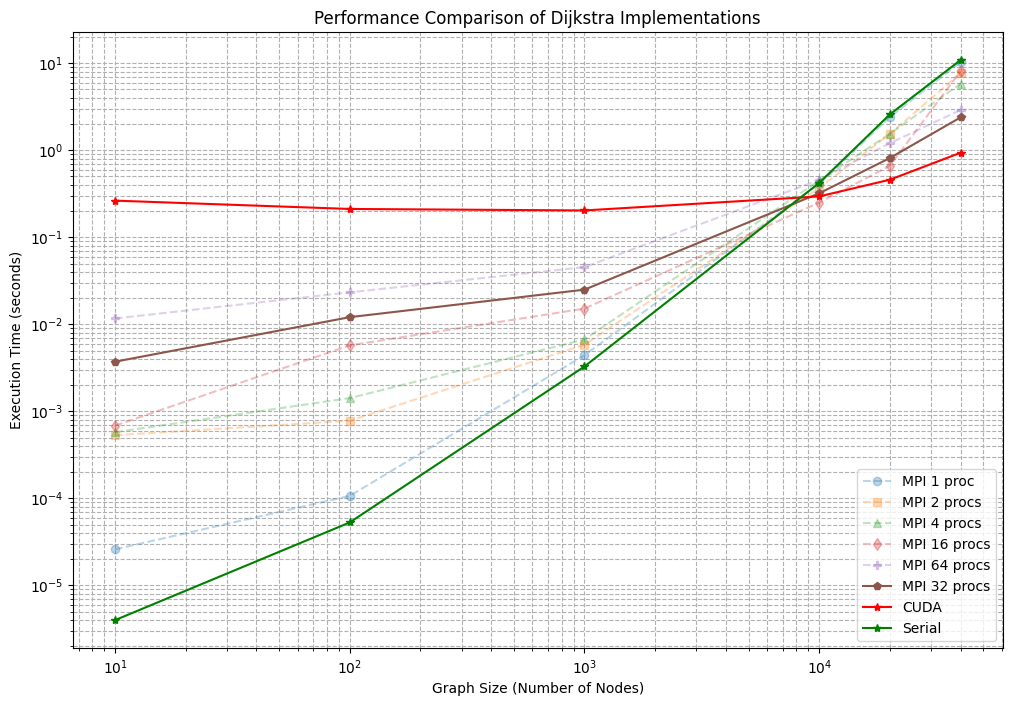}
    \caption{Performance (log scale)}
\end{figure}

\columnbreak
\subsubsectionstar{Performance analysis}
For small-sized graphs, both parallel implementations (MPI and CUDA) perform worse than the serial implementation due to communication overhead and synchronization overhead. However, as the graph size increases, the serial performance\footnote{The time complexity of serial implementation: $O(n^2)$} drops rapidly. For large graphs, both MPI parallelism and CUDA parallelism show excellent performance. MPI parallelism brings about 5 times of performance improvement, while CUDA parallelism brings more than 10 times of performance improvement.

\subsection{Dense and Sparse Graphs}
\label{sec:Dense and Sparse Graphs}
A graph with the number of edges equal to the square of the number of nodes is called a dense graph, which means that most of the nodes are reachable. My implementation uses the adjacency matrix, the density of the graph has little impact on the results, because in the adjacency matrix, unreachable edges are represented as $\infty$. The calculation logic and cost of unreachable edges are similar to weighted reachable edges. To verify my above conclusion, I compared the performance of sparse graphs and dense graphs under the same nodes.

\begin{table}[H]
\centering
\resizebox{\columnwidth}{!}{%
\begin{tabular}{|c|c|c|c|c|}
\hline
\# of Nodes & \# of Edges & Serial time & MPI (32 procs) time &CUDA time \\
\hline
10 & 30 & 0.000004&0.003724&0.330707 \\
10 & 45 & 0.000004&0.003756&0.254294 \\
\hline
100 & 300 & 0.000074&0.012098&0.251250 \\
100 & 4950 & 0.000057&0.011727&0.232588 \\
\hline
1000 & 3000 & 0.003287&0.024123&0.228707 \\
1000 & 499500 & 0.003837&0.024084&0.215282 \\
\hline
2000 & 6000 & 0.012743&0.043465&0.220302 \\
2000 & 1899500 & 0.017153&0.045017&0.215418 \\
\hline
\end{tabular}
}
\caption{Dense Graphs}
\end{table}

Pay attention to the time in each small square. Under the same \# of nodes, the processing time of sparse\footnote{Number of Nodes : Number of Edges = 1 : 3} or dense\footnote{Number of Nodes : Number of Edges $\approx$ n : $\frac{n^2}{2}$} graphs is similar.

\subsection{MPI parallelism Details}
\label{sec:MPI parallelism Details}
\begin{figure}[h]
    \centering
    \includegraphics[width=0.9\columnwidth]{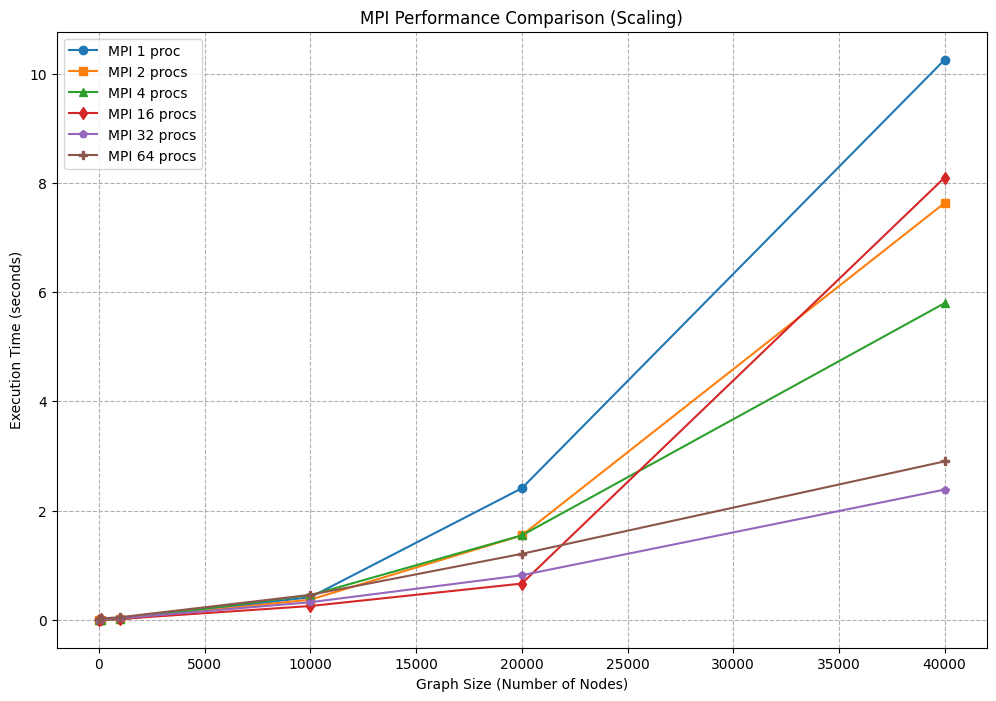}
    \caption{Performance (linear scale)}
\end{figure}

\begin{figure}[h]
    \centering
    \includegraphics[width=0.9\columnwidth]{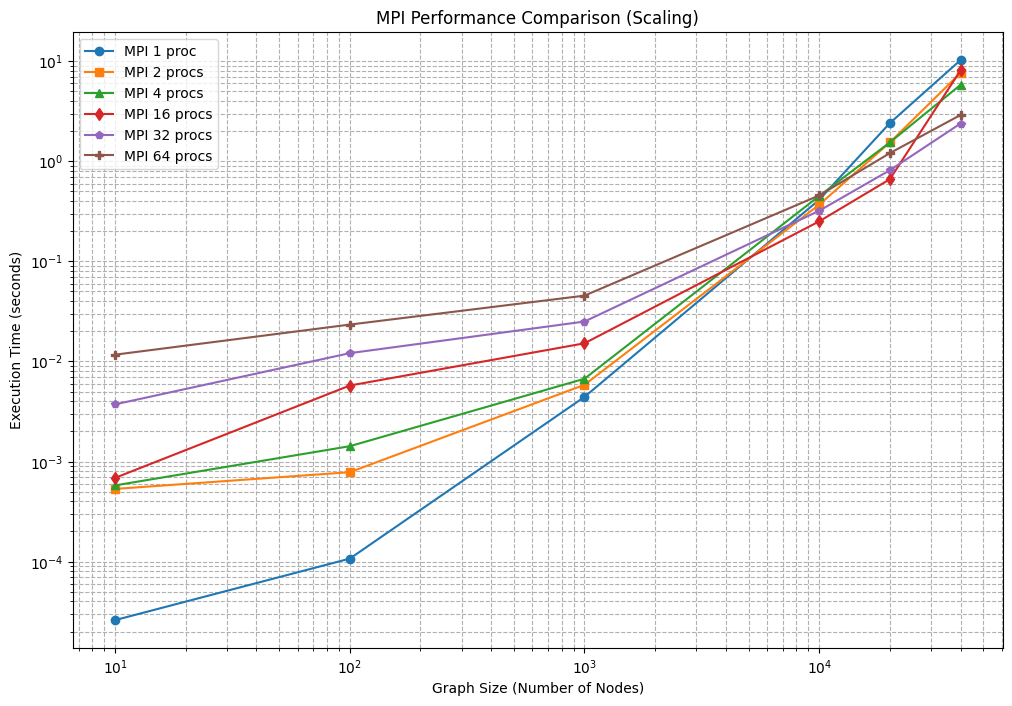}
    \caption{Performance (log scale)}
\end{figure}

The performance of 32 procs exceeds that of 64 procs. My analysis is that communication overhead becomes the main limitation of performance. When the number of MPI processes increases, the amount of data processed by each process decreases, but the data exchange and synchronization requirements between processes increase. This is a point that needs to be optimized in my MPI parallel implementation. I need to design a higher performance MPI to reduce the synchronization overhead between processes.

\subsubsectionstar{Scaling\footnote{The current implementation cannot perform efficient weak scaling because if 1 proc is used to process at large scales (e.g. 10000), 64 procs need (640000) and the graph file is difficult to generate. Small-scale graph scaling cannot reflect performance}}

\begin{table}[H]
\centering
\begin{tabular}{|l|c|r|}
\hline
procs & time (s)  & scaling efficiency (\%) \\
\hline
1     & 10.28     &   \\
2     & 7.67      & 67.01\\
4     & 5.88      & 43.71\\
16    & 8.05      & 7.98 \\
32    & 2.36      & 13.61\\
64    & 2.90      & 5.53 \\
\hline
\end{tabular}
\caption{Strong scaling - $4\times10^4$ Nodes}
\label{tab:strong_scaling}
\end{table}

\begin{figure}[H]
\centering
\includegraphics[width=0.8\columnwidth]{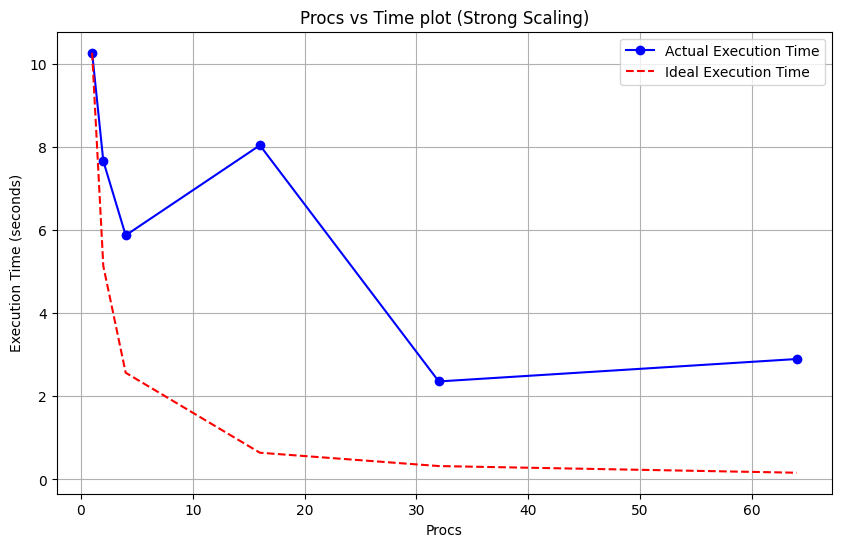}
\caption{Strong Scaling plot}
\end{figure}

%
\section{Conclusion - Challenges and Obstacles}
Paralleling dijkstra's algorithm is challenging. Compared with Particle Simulation, the classic implementation of dijkstra's algorithm has serial logic (continuously finding the node closest to the starting point among the currently unprocessed nodes). Even if it is parallel, it requires frequent synchronization operations, which makes its parallel effect not particularly significant. My parallel version of MPI achieves \textbf{5x} the performance of the serial benchmark on large scale graphs, parallel version of CUDA achieves \textbf{10x} the performance of the serial benchmark on large scale graphs. This improvement is not particularly good in high-performance parallel optimization. I analyze the reasons as follows.
\begin{enumerate}
    \item The "inherent shortcomings" of dijkstra's algorithm parallelism.
    \item Code design makes synchronization overhead too large, which is deeply reflected in MPI parallelism. My MPI parallel synchronization operations are not granular enough. If I design the synchronization  better, I may be able to achieve better performance improvements.
    \item Adjacency matrix as data structure. Although the adjacency matrix as a data structure allows the implementations have the same performance when processing sparse graphs and dense graphs, it requires allocating \textit{n$\times$n} memory, which wastes a lot of overhead when processing sparse graphs.
\end{enumerate}
\bibliographystyle{IEEEtran}
\bibliography{references}

\begin{thebibliography}{1}
\providecommand{\url}[1]{#1}
\csname url@samestyle\endcsname
\providecommand{\newblock}{\relax}
\providecommand{\bibinfo}[2]{#2}
\providecommand{\BIBentrySTDinterwordspacing}{\spaceskip=0pt\relax}
\providecommand{\BIBentryALTinterwordstretchfactor}{4}
\providecommand{\BIBentryALTinterwordspacing}{\spaceskip=\fontdimen2\font plus
\BIBentryALTinterwordstretchfactor\fontdimen3\font minus \fontdimen4\font\relax}
\providecommand{\BIBforeignlanguage}[2]{{%
\expandafter\ifx\csname l@#1\endcsname\relax
\typeout{** WARNING: IEEEtran.bst: No hyphenation pattern has been}%
\typeout{** loaded for the language `#1'. Using the pattern for}%
\typeout{** the default language instead.}%
\else
\language=\csname l@#1\endcsname
\fi
#2}}
\providecommand{\BIBdecl}{\relax}
\BIBdecl

\bibitem{DIJKSTRA1959}
\BIBentryALTinterwordspacing
E.~DIJKSTRA, ``A note on two problems in connexion with graphs.'' \emph{Numerische Mathematik}, vol.~1, pp. 269--271, 1959. [Online]. Available: \url{http://eudml.org/doc/131436}
\BIBentrySTDinterwordspacing

\end{thebibliography}

\end{document}